\documentstyle[prl,twocolumn,aps,epsf]{revtex}
\newcommand{\be}{\begin{equation}}
\newcommand{\ee}{\end{equation}}
\newcommand{\bea}{\begin{eqnarray}}
\newcommand{\eea}{\end{eqnarray}}

\begin{document} 

\twocolumn[\hsize\textwidth\columnwidth\hsize\csname@twocolumnfalse\endcsname

\title {On the relative positions of the $2\Delta$ peaks in 
Raman and tunneling spectra of $d-$wave superconductors.}
\author{Andrey V. Chubukov, Nathan Gemelke, and Ar. Abanov}
\address{
Department of Physics, University of Wisconsin, Madison, WI 53706}
\date{\today}
\draft
\maketitle 
\begin{abstract}
We study $B_{1g}$ Raman intensity $R(\Omega)$ and the density of states
$N(\omega)$ in isotropic $2D$ $d-$wave superconductors. 
For an
ideal gas, $R(\Omega)$ and $N(\omega)$ have sharp peaks at $\Omega =2\Delta$
and $\omega =\Delta$, respectively,
 where $\Delta$ is the maximum value of the gap.
 We study how the peak positions are affected by the
fermionic damping due to impurity scattering.
 We  show that while the damping generally shifts
the peak positions to larger frequencies,  the peak in $R(\Omega)$
still occurs at almost 
twice the peak position in $N(\omega)$ and therefore cannot
account for the  experimentally observed downturn shift of
the peak frequency in $R(\Omega)$ in underdoped cuprates compared to
twice that in $N(\omega)$. We also discuss how the fermionic
damping affects the dynamical spin susceptibility.
\end{abstract}

]
\narrowtext

The unusual physical properties of cuprate superconductors have continued
to be of high interest to condensed matter physicists for more than a
decade.  In recent years, a lot of attention was devoted to the study of
collective bosonic excitations in the superconducting and pseudogap
phases\cite{neutrons,neutrons2,Blu98}.  The most notable experimental
observation is the discovery of the resonance mode in the dynamical spin
susceptibility~\cite{neutrons,neutrons2}.  This mode is centered at the
antiferromagnetic momentum $Q=(\pi,\pi)$, and physically reflects the fact
that near antiferromagnetic instability, collective spin excitations in a
d-wave superconductor are undamped, propagating spin waves at energies
smaller than $2\Delta$~\cite{mazin,schr-shen,ding,ac,morr}. 
Less attention is devoted to the study of possible  resonance bosonic
excitations at zero momentum transfer. These excitations are probed by
Raman scattering which generally measures the imaginary part of the fully
renormalized 
particle-hole susceptibility at vanishingly small incoming momentum, 
weighted with Raman form factors which depend on the scattering geometry
~\cite{Abrikosov,Klein84}. 
The experiments relevant to our discussion were performed in $B_{1g}$
geometry where the Raman form 
factors are the largest for fermionic momenta near $(0,\pi)$ and symmetry
related points where the $d_{x^2-y^2}$ gap
$\Delta (k)$ is near its maximum
$\Delta$~\cite{Shastry,Devereaux}. 

In a BCS theory for a 
$d-$wave superconductor  $B_{1g}$ Raman intensity $R(\Omega)$ 
logarithmically diverges at
$2\Delta$ and rapidly, as $\omega^3$, decreases at smaller 
frequencies~\cite{Devereaux}.  Experimental data for overdoped $Bi2212$ are
qualitatively consistent with this behavior~\cite{Blu98,Hackl96}.
Furthermore, the $2\Delta$ extracted from $R(\Omega)$ is almost exactly
twice the gap extracted from SIN tunneling data, which measure a single
particle density of states (DOS) $N(\omega)$~\cite{tunn}. With
underdoping, however, the peak frequency in $R(\Omega)$ progressively
deviates down from the $2\Delta$ extracted from the
tunneling experiments~\cite{Blu98}.
    
Blumberg, Morr and one of us (CBM)~\cite{cbm}  attributed
this deviation to a final state interaction between scattered quasiparticles. 
They argued that the magnetically mediated final state interaction 
in $B_{1g}$ geometry is attractive and  gives rise to a
pseudo-resonance in $R(\Omega)$ at a frequency $\Omega_{res}$, which with
underdoping progressively deviates down from $2\Delta$.

An alternative to the resonance mode scenario is one in which the final
state interaction is irrelevant, and the discrepancy between Raman and
tunneling data is due to fermionic incoherence, which generally 
shifts the positions of both the Raman peak and the peak in the DOS. In this
paper, we show that the  shifts in the peak positions of $R(\Omega)$ and
$N(\omega)$ due to fermionic damping are
{\it correlated} such that without final state interaction, 
the peak in $R(\Omega)$ is still located at almost 
exactly twice the peak frequency in $N(\omega)$. 
This result implies that the
experimentally observed  relative
downturn deviation of the peak in the Raman intensity 
{\it cannot} be explained by purely fermionic
self-energy effects, and leaves the resonance mode scenario as the most 
probable one. 

We begin with the general expressions for $R(\Omega)$ and $N(\omega)$
in a superconductor. 
The DOS  is the imaginary part of the local normal
 quasiparticle Green's function, and the 
 Raman intensity without final state interaction 
is the imaginary part of the fermionic polarization 
bubble with a zero momentum transfer, weighted with Raman vertices~
\cite{Schrieffer,m-z}. We have, up to an overall factor 
\begin{eqnarray}
&& R(\Omega) = Im~\int d k~ d \omega ~
~V^2_{B_{1g}}({\bf k})~(G_{sc} (k,\omega_+) ~
G_{sc} (k,\omega_-) + \nonumber \\   
&&F(k,\omega_+)~ F(k, \omega_-));
~N(\omega) =  Im~ \int d k~ G_{sc} (k,\omega)
\label{R_sup}
\end{eqnarray} 
Here, $V_{B_{1g}}({\bf k}) \propto \cos k_x - \cos k_y$ is the $B_{1g}$
Raman vertex, $\omega_\pm = \omega \pm \Omega/2$, and $G_{sc} (k, \omega)$
and $F(k,\omega)$ are normal and anomalous quasiparticle Green's functions
given by
\begin{eqnarray}
G_{sc} (k, \omega) &=& G^{-1}_n (-k, -\omega)/
(G^{-1}_n (k, \omega)~G^{-1}_n (-k, -\omega) + \Delta^2_k)
\nonumber \\
F (k, \omega) &=& 
i \Delta_k/(G^{-1}_n (k, \omega)~G^{-1}_n (-k, -\omega) + 
\Delta^2_k) \ .
\label{GF}
\end{eqnarray}
Here,
 $G^{-1}_n
(k,\omega) = \Sigma({\bf k},\omega) - \epsilon_k$,  where $\Sigma ({\bf
k},\omega)$ is the fermionic self-energy (which also absorbs a bare 
$\omega$ term), and 
 $\Delta_k$ 
is the superconducting gap, which for a $d_{x^2 -y^2}$ superconductor
behaves as
 $\Delta_k \propto \cos k_x - \cos k_y$.

In an ideal gas, the fermionic self-energy is absent (i.e., $\Sigma ({\bf
k},\omega) = \omega$). To simplify the discussion, we assume that the
Fermi surface is circular, for which both $R(\Omega)$ and
$N(\omega)$ can be evaluated exactly~\cite{Devereaux}.
Substituting the momentum integration by integration
over $d\epsilon_k$, and approximating the $k-$dependences of
$V_{B_{1g}}({\bf k})$ and
$\Delta_k$  by $\cos \Theta$, we obtain for ${\bar \Omega},~{\bar \omega} <1$
\begin{equation}
R({\bar\Omega}) = 
%{\bar \Omega}^3
%\int_0^{\pi/2}\frac{d \Theta~ cos^4
%\Theta}{\sqrt{1- {\bar \Omega}^2 \cos^2 \Theta}} =
 \frac{3\pi}{16} 
~{\bar \Omega}^3~F(\frac{1}{2},\frac{5}{2},3,{\bar \Omega}^2),~
N({\bar \omega}) 
%&=& \frac{2}{\pi} {\bar \omega}~ 
%\int_0^{\pi/2} \frac{d \Theta}{\sqrt{1 - {\bar \omega}^2 \cos^2 \Theta}} = 
= \frac{2}{\pi}~{\bar \omega}~K({\bar \omega}^2),\nonumber
\end{equation}
and for  ${\bar \Omega},~{\bar \omega} >1$
\begin{equation}
R({\bar\Omega}) = 
%{\bar \Omega}^{-2}
%\int_0^{\pi/2} \frac{d \Theta~ cos^4
%\Theta}{\sqrt{1- {\bar \Omega}^{-2} \cos^2 \Theta}} = 
\frac{3\pi}{16{\bar \Omega}^{2}}~F(\frac{1}{2},\frac{5}{2},3,\frac{1}{{\bar
\Omega}^{2}}),
%;\nonumber\\
N({\bar\omega}) 
%&=& \frac{2}{\pi} 
%\int_0^{\pi/2} \frac{d \Theta}{\sqrt{1 - {\bar \omega}^{-2} \cos^2 \Theta}} = 
= \frac{2}{\pi}~K\left(\frac{1}{{\bar \omega}^{2}}\right),
\label{idgas}
\end{equation}
where  ${\bar \omega} =
\omega/\Delta$, ${\bar \Omega} =
\Omega/(2\Delta)$, and  $F(a,b,c,x)$ and $K(x)$ are
 hypergeometric and elliptical functions, respectively.
\begin{figure}
  \centerline{\epsfxsize=3.0in \epsfysize=3.0in 
  \epsffile{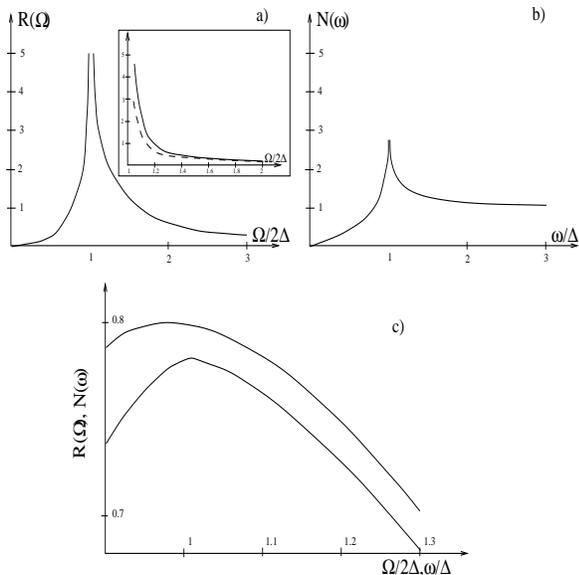}}
\caption{The behavior of the Raman intensity $R(\Omega)$ and the DOS
$N(\omega)$ in $d-$wave superconductors. a), b) Fermi-gas results ($\gamma =0$)
Solid lines -$d-$wave
results, dashed lines -$s-$wave results shown for comparison.
 The insert shows the behavior of
$R(\Omega)$ close to the threshold frequency $2 \Delta$.
 c)
The results for $\gamma = 0.2 \Delta$ (Eq. (\protect\ref{resd}))
 The normalized frequencies are
 ${\bar \omega} = \omega/\Delta$ and ${\bar \Omega} = \Omega/(2\Delta)$.
Observe that for $d-$wave superconductors, the peak in the Raman intensity is
located at 
larger normalized frequency than the peak in the density of states.} 
\label{fig1} 
\end{figure} 
The results for $R(\Omega)$ and
$N(\omega)$ are plotted in Fig. ~\ref{fig1}a,b. 
At the smallest frequencies, 
$R(\Omega) \propto
\Omega^3$ and $N(\omega) \propto \omega$. At larger frequencies, 
the Raman intensity and the DOS  diverge logarithmically 
at $\omega = \Delta$ and $\Omega = 2\Delta$, respectively.
At even larger frequencies, the momentum dependence of the gap
progressively becomes less relevant, 
and both $R(\Omega)$ and
$N(\omega)$ acquire the same forms
 as in $s-$wave superconductors:
$R({\bar\Omega}) \propto 1/( {\bar \Omega} \sqrt{{\bar \Omega}^2 -1}),~
N({\bar\omega}) = {\bar \omega}/\sqrt{{\bar \omega}^2 -1}.$ Observe that
$R(\Omega)$ crosses over to the $s-$wave behavior immediately above $2\Delta$.
(see inset in Fig. ~\ref{fig1}a).

Our goal is to study how the  peak positions and, more generally, 
the functional forms of $R(\Omega)$ and $N(\omega)$ 
are affected by the fermionic self-energy. In general, the 
fermionic self-energy comes from various sources, 
and at least part of it, associated with the scattering by the same
bosonic excitations which give rise to superconductivity, has to be
determined fully self-consistently from the Eliashberg-type
equations~\cite{ac}. 
In this paper, we assume for simplicity that
the primary source for the fermionic damping
is impurity scattering. We 
consider the fermionic self-energy in the self-consistent $T-$matrix formalism~\cite{peter,dk}
and neglect subtle 2D effects beyond $T-$matrix 
approximation~\cite{lee}. 
Hirshfeld, Wolfle and Einzel~\cite{peter} demonstrated that for a $d-$wave
superconductor with a particle-hole symmetry and $k-$independent scattering
potential,
$\Sigma (k, \omega) = \omega + i~\gamma_{\omega}~ sign (\omega)$, where
$\gamma_{|\omega|}$ is a solution of the self-consistent 
equation~\cite{peter,dk}
\begin{equation}
\gamma_{|\omega|} = ~\gamma \frac{g_0 (|\omega|)}{c^2 + g^2_0 (|\omega|)}
\label{sc}
\end{equation}
Here $\gamma$ is proportional to the impurity concentration, $c$ is the
cotangent of the scattering phase shift, and $g_0 (|\omega|) = (i/\pi N_F)
\sum_k G_{sc} (k, \omega)~sign (\omega)$, where $N_F$ is the normal state 
DOS at the Fermi
surface. The self-consistency of (\ref{sc})
is in the fact that $G_{sc} (k, \omega)$ given by
(\ref{GF}) by itself depends on $\gamma (\omega)$ through $\Sigma
(\omega)$. Physically this 
means that 
the fermionic self-energy due to impurity scattering is by itself affected by
a superconductivity.

For a circular Fermi surface, the momentum integral over $G(k,
\omega)$ can be performed exactly and yields
$g_0 (|\omega|) =  (2/\pi) K
(\Delta^2/\Sigma^2(\omega))$. 
In the normal state, $g_0 (\omega) = 1$,
 and $\gamma_{|\omega|}$
reduces to a constant 
$\gamma_{|\omega|} =  \gamma /(1+
c^2)$. In a superconducting state, however, 
$\gamma_{|\omega|}$ is complex and
frequency dependent.
Still, one can easily demonstrate that $\gamma_{|\omega|}$ remains finite
 for all frequencies.
Below we will need $\gamma_0$ and 
 $\gamma_\Delta$.
For unitary
scattering (c=0), we solved (\ref{sc}) at these frequencies 
and for $\gamma \ll \Delta$
obtained with logarithmical accuracy
$\gamma_\Delta = (2/\pi) \gamma 
\log (\Delta/\gamma)$ and 
$\gamma_0 = (\pi \gamma \Delta/\log(\Delta/\gamma))^{1/2}$.

We now proceed with the calculations of the DOS and Raman intensity. 
As the self-energy is
$k$-independent, we can use the same trick as in earlier
studies~\cite{ac,cbm,cm}, and first integrate over momenta in (\ref{R_sup}).
  Substituting the momentum integration by the
integration over $\epsilon_k$ and evaluating the integrals,
 we obtain~\cite{cbm}
\begin{eqnarray}
&&N(\omega)= Im \int_0^{\pi/2} d\Theta~
\frac{\Sigma (\omega)}{D(\omega)};~
R(\Omega) = - Re \int_0^{\pi/2} d\Theta \cos^2 \Theta~\nonumber \\
&& \times \int _{-\infty }^{\infty }d \omega \frac {(\Sigma_+ -\Sigma_-)^2 + 
(D_+-D_-)^2}{4D_+D_- (D_++D_-)}.
\label{chi}
\end{eqnarray}
Here $\omega_{\pm }= \omega \pm \Omega/2$,
$\Sigma_\pm = \Sigma (\omega_\pm),  ~D_\pm = D (\omega_\pm)$, and 
$D(\omega) = 
\sqrt {\Delta^2 \cos^2 \Theta  - \Sigma ^2({\omega })}$.

As a warm up, consider the limit of small frequencies.
Substituting $\Sigma (\omega) = \omega + i \gamma_0 sign (\omega)$ into
(\ref{chi}) and expanding in frequency, we obtained
\begin{eqnarray}
R({\bar \Omega}) &=& {\bar \Omega}~{\bar \gamma}_0^2 \log{1/{\bar \gamma}_0} +
O({\bar\Omega}^3)
\nonumber \\
N({\bar \omega}) &=& N(0) + {\bar \omega}~(1 - \frac{2}{\pi}
tan^{-1}{\frac{{\bar\gamma}_0}{\bar \omega}}) - 
\frac{2{\bar \gamma}_0}{\pi} \log{\sqrt{1 +
(\frac{\bar \omega}{{\bar\gamma}_0})^2}}
\label{smo}
\end{eqnarray}
\begin{figure}
  \centerline{\epsfxsize=3.0in \epsfysize=3.0in 
  \epsffile{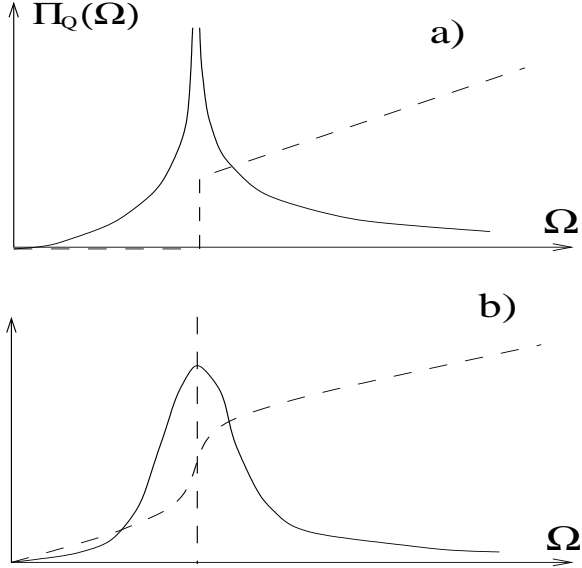}}
\caption{The behavior of the spin polarization operator in the Fermi gas (a)
and at a finite $\gamma$ (b). Solid line - $Re \Pi_Q (\Omega)$, dashed line -
$Im \Pi_Q (\Omega)$}
\label{fig2} 
\end{figure} 
where $N(0) = (2/\pi) {\bar \gamma}_0 \log{1/{\bar \gamma}_0}$,
and ${\bar \gamma}_0 = \gamma_0/\Delta$.  At ${\bar \omega}
\ll 1, ~N({\bar \omega}) = N(0) + {\bar \omega}^2/(\pi {\bar \gamma}_0)$.

We see that fermionic damping (i) yields a linear frequency dependence
of $R(\Omega)$, and (ii) 
yields a finite DOS at zero frequency, and the 
quadratic frequency dependence of $N({\bar \omega})$ above $N(0)$.
Both of these results agree with earlier studies~\cite{peter,dk,Gorkov}.

We now consider $R(\Omega)$ and 
at $\Omega \approx 2\Delta$ and $\omega
\approx \Delta$ where the Raman intensity and the DOS diverge
in a Fermi gas.
Simple estimates show that near $2\Delta$, the integral in $R(\Omega)$ 
is dominated by
$D_{\pm } \ll 1$ i.e. $\omega_\pm \approx \pm \Delta$.  Substituting
$\Sigma (\omega \approx \Delta) = \omega + i\gamma_\Delta~sign (\omega)$ 
into (\ref{chi}) and expanding  the integrands to
first order in $\Omega - 2\Delta$ and $\omega - \Delta$ respectively,
 we obtain after lengthy but straightforward calculations
\begin{eqnarray}
R ({\bar \Omega}) &=&  
%\frac {1}{2\sqrt {\bar \gamma}_\Delta}~ \int_0^{\pi/2} d\Theta~ \cos^2 \Theta 
%\left(~\frac {\sqrt {1+\alpha^2_{\bar\Omega}} + \alpha_{\bar\Omega}}{1+
%\alpha^2_{\bar\Omega}}\right)^{\frac {1}{2}}~ \nonumber \\
%&& = 
\frac{\pi}{4 \sqrt{2}}~Re\left[\frac{F \left(\frac{1}{2},\frac{1}{2},2,
\frac{-2}{\bar \Omega -1 + 
i {\bar \gamma}_\Delta}\right)}{\sqrt{\bar \Omega -1 + 
i {\bar \gamma}_\Delta}} \right];
\nonumber \\
N ({\bar \omega})  &=&  
%\frac {1}{\pi\sqrt {\bar \gamma}_\Delta}~\int_0^{\pi/2} d\Theta~ 
%~\left(\frac {\sqrt {1+\alpha^2_{\bar\omega}} + 
%\alpha_{\bar\omega}}{1+
%\alpha^2_{\bar\omega}}\right)^{\frac {1}{2}}
%~ \nonumber \\ 
%&&= 
\frac{\sqrt{2}}{\pi}~
Re\left[\frac{K \left(\frac{-2}{\bar \omega -1 + 
i \bar\gamma_\Delta}\right)}{\sqrt{\bar \omega -1 + 
i \bar\gamma_\Delta}}\right]
\label{resd}
\end{eqnarray}
Here 
%$\alpha_x = (x-1 + 2 \sin^2 \Theta)/({\bar \gamma}_\Delta)$ and
$\bar \gamma_\Delta =
\gamma_\Delta/\Delta$. 

We now analyse these results. 
Near ${\bar \Omega} = {\bar \omega} =1$, we find from (\ref{resd})
%\begin{equation}
$R({\bar \Omega}) \approx (-1/8) 
\log[({\bar \Omega} -1)^2 + {\bar\gamma}_\Delta^2],~
N({\bar \omega}) \approx (-1/4\pi)   
\log[({\bar \omega} -1)^2 + {\bar\gamma}_\Delta^2]$.
We see that the  logarithmical divergencies are
 cut by the fermionic damping, but, to a logarithmical accuracy, the peaks
remain at the same positions as in a Fermi gas.
In other words, fermionic damping gives rise to a broadening of the 
peaks in the Raman intensity and the DOS, but still, the peak in $R(\Omega)$
 is
located at twice the peak frequency of the DOS. 
%This last result can be easily
%understood as the logarithmical singularity comes from the momentum
%range where the gap is near maximum, and the Raman vertex $V^2_{B_{1g}} = 
%\cos^2 \Theta \approx 1$. Without Raman vertex, the functional forms of $R({\bar
%\Omega})$ and $N({\bar \omega})$ are equivalent (see (\ref{resd})),
%and hence the peak in
%$R(\Omega)$ should be exactly at twice the peak frequency in $N(\omega)$. 

Calculations beyond the logarithmical accuracy  show that 
the peak positions do shift to higher
frequencies but the relative shift is 
opposite to the one detected in the experiments: the peak in $R({\bar \Omega})$
shifts to higher frequencies for arbitrary $\gamma_\Delta$ 
(${\bar \Omega}_{peak} -1 \propto {\bar \gamma}_\Delta^2$ for ${\bar
\gamma}_\Delta \ll 1$),
while the peak in $N({\bar \omega})$ shifts to high frequencies only if the 
damping exceeds a threshold value of ${\bar \gamma}_\Delta \approx 0.77$. 
Obviously,
the magnitude of the shift in $N(\omega)$ is smaller than that in
$R(\Omega)$. 
The behavior of $N({\bar \omega})$ and $R({\bar \Omega})$ is illustrated in
Fig.~\ref{fig1}c.
 
Note in passing that  for $s-$wave superconductors, the
% results  for $R({\bar \Omega})$ and $N({\bar\omega})$ 
the same calculations which lead to  
%are obtained from 
(\ref{resd}) yield
%by eliminating the $\sin^2 \Theta$ from $\alpha_x$. This yields 
%\begin{eqnarray}
\begin{equation}
%R({\bar \Omega}) \sim
% &=&  
%\frac {\pi}{8\sqrt {\bar \gamma}_\Delta} 
R(x), N(x) \sim \left(\frac {\sqrt {1+\alpha^2_x} + \alpha_x}{(1+
\alpha^2_x){\bar \gamma}_\Delta}\right)^{\frac {1}{2}}
%
%\left(\frac {\sqrt {1+\alpha^2_{\bar\Omega}} + \alpha_{\bar\Omega}}{(1+
%\alpha^2_{\bar\Omega}){\bar \gamma}_\Delta}\right)^{\frac {1}{2}};~
% \nonumber \\
%N({\bar \omega}) \sim
%&=& 
%\frac{1}{2 \sqrt{{\bar \gamma}_\Delta}}~
%\left(\frac {\sqrt {1+\alpha^2_{\bar\omega}} + 
%\alpha_{\bar\omega}}{(1+\alpha^2_{\bar\omega}){\bar \gamma}_\Delta}\right)^{\frac {1}{2}}
%\nonumber
%\label{ress}
%\end{eqnarray}
\end{equation}
where $\alpha_x = (x-1)/({\bar \gamma}_\Delta)$ and 
$x = {\bar \Omega}$ for $R(x)$ and $x= {\bar \omega}$ for $N(x)$.
Again, the divergencies at
$\bar \Omega = \bar \omega =1$ are gone and substituted by the
peaks at  higher frequencies for which 
$\alpha_x =  \frac{1}{2\sqrt 3}$.
At the same time, 
%we  see from (\ref{ress}) that 
the functional forms of $R ({\bar \Omega})$ and
$N({\bar \omega})$ are {\it identical}. This implies that in a dirty $s-$wave
superconductor, the peak in
the Raman intensity is also located
exactly at twice the peak frequency in $N(\omega)$, though both peak
positions shift from the Fermi gas values.

For completeness, we also discuss how fermionic damping affects the spin
polarization operator at the 
antiferromagnetic momentum $Q$. The form of this
polarization operator is relevant for the interpretation of neutron
scattering and ARPES data~\cite{ding,ac,cm}.
 
The spin polarization operator is 
related to the dynamical structure factor as
$S_Q (\bar \Omega) \propto Im (\xi^{-2} - \Pi_Q (\bar
\Omega))^{-1}$ where $\xi$ is the magnetic correlation length~\cite{ac}. It is
formally 
given by the same set of particle-hole
bubbles made of normal and anomalous Green's functions as the Raman intensity,
but differs from $R(\Omega)$ in two aspects.
First, the antiferromagnetic spin polarization is a finite momentum probe, 
and the contribution to low frequency 
$Im \, \Pi_Q (\Omega)$ only comes from the momentum range in
the Brillouin zone where both particles in the bubble are near the Fermi 
surface hot spots. Near hot spots,
the superconducting gap is finite and close to $\Delta$.
In other words, the regions near the nodes of the
$d_{x^2-y^2}$ gap do not contribute to the dynamical spin susceptibility near
$Q$. Second,  the vertices for $\Pi_Q (\Omega)$
contain Pauli matrices. For the anomalous 
$FF$ term, the summation over spin projections
yields an extra factor $-1$ compared to the Raman bubble.
Performing the momentum integration in the $GG$ and $FF$ bubbles
in the same
way as before, and using the fact that 
$\Delta_k \Delta_{k+Q} = -\Delta^2$, 
we obtain~\cite{ac}
\begin{equation}  
\Pi_Q (\Omega) = i 
\int _{-\infty}^{\infty } d \omega 
\frac { \Sigma _+ \Sigma _- + D_+ D_-  -\Delta^2}{2 D_+D_-}
\label{Pi}
\end{equation}
The overall factor is chosen such that in the normal state, $\Pi_Q (\Omega) = i
|\omega|$.

For an ideal
gas, the frequency integration in (\ref{Pi}) yields 
$Im~\Pi_Q ({\bar\Omega}) =0$ and  $Re~\Pi_Q ({\bar \Omega}) 
\propto {\bar \Omega}^2$ at ${\bar \Omega} = \Omega/(2\Delta) <1$.
For large enough $\xi$, this behavior of $\Pi_Q (\Omega)$ gives rise
to a resonant peak 
in $S_Q (\bar \Omega)$ at
a frequency where  $Re \, \Pi_Q (\Omega) = \xi^{-2}$~\cite{mazin,ac}.
At ${\bar \Omega} =1$, $Im \, \Pi_Q (\Omega)$ jumps to a finite value, and
$Re \, \Pi_Q(\Omega)$ diverges logarithmically~\cite{mazin}. This behavior
is shown in
Fig.~\ref{fig2}a.

 Substituting $\Sigma (|\omega| \approx \Delta)$ into (\ref{Pi}) 
and performing the same calculations as before, we obtain near ${\bar
\Omega} =1$ and to first order in $\bar\gamma_\Delta$
\begin{eqnarray}
Im \, \Pi _Q ({\bar \Omega }) &=& \frac{\Delta \pi}{2}
 (1 + \frac {2}{\pi }\arcsin {\frac {\alpha_{\bar \Omega}}{\sqrt
{\alpha^2_{\bar \Omega} +1}}})\nonumber
\\
Re \, \Pi _Q ({\bar \Omega }) &=& \Delta
\left( \log{\frac{1}{\bar\gamma}_\Delta} - \Psi (\alpha_{\bar \Omega})\right)
\end{eqnarray}
where, as before, $\alpha_{\bar\Omega} = (\bar \Omega-1)/(\bar\gamma_\Delta)$, 
and in the limits of small and large $\alpha$, $\Psi (\alpha)$ behaves as
$\Psi (\alpha \ll 1) = \alpha^2 /2$, and $\Psi (|\alpha| \rightarrow
\infty) = \log |\alpha|$.

We see, similar to what we found for the Raman intensity and the
DOS, the inclusion of the fermionic damping eliminates the singularities
in the spin polarization operator: $Im \, \Pi _Q$ changes continuously
through $\bar {\Omega } =1$, and $Re \, \Pi _Q$ is peaked but does not
diverge at $\bar \Omega = 1$  (to first order in ${\bar \gamma}_\Delta$,
 the peak does not shift to higher frequencies). This behavior is shown in
Fig.~\ref{fig2}b.

At small frequencies, the
%straightforward 
expansion in $\bar \Omega$ in (\ref{Pi}) yields 
\begin{equation}
\Pi _Q ({\bar \Omega }) = \frac{\pi}{2} 
{\bar \Omega}^2  + 2i {\bar\gamma_0}^2~|{\bar
\Omega}|.
\end{equation}
Again, similar to the result for $R(\Omega)$,
the inclusion of a finite fermionic damping yields a nonzero $Im \, \Pi_Q
(\Omega)$
down to the lowest frequencies.
This result implies that in the presence of  impurity scattering, 
the resonance peak in $S_Q (\Omega)$ has a finite width.
 This effect may
account for the width of the resonance neutron peak near optimal doping.
We, however, do not believe that fermionic damping is responsible for the
increase of the peak width with underdoping - this last effect is likely to
be caused by the frequency dependence of $\Delta$ associated with the
pseudogap effects, which we do not consider here~\cite{acf}. 

To summarize, we have considered in this paper a simple model form of the 
electronic damping and analysed how it affects the forms of the Raman
intensity, the DOS, and the spin polarization operator at the
antiferromagnetic momentum. We found that a finite damping eliminates
artificial divergencies found in a Fermi-gas consideration. Still,
however, without final state interaction, the peak in $R(\Omega)$ occurs
at or beyond twice the peak frequency for the DOS, in contradiction with
the experimental observations. This negative result implies that fermionic
damping alone cannot account for the data, and supports the
explanation of the downturn shift of the Raman peak with underdoping
in terms of a mid-gap pseuo-resonance mode in the Raman
intensity~\cite{cbm}.  We also found that fermionic damping gives rise to a finite
width of the resonance neutron peak.

It is our pleasure to thank G. Blumberg, A. Finkel'stein, D. Morr,
and  M. Norman for useful conversations.  The research was supported by NSF
DMR-9629839.

\end{document}